\documentclass[prb,aps,amsmath,amssymb,twocolumn,groupedaddress,floats,showpacs,final,reprint]{revtex4-1}
\usepackage{graphicx}
\usepackage{amssymb,amsmath}
\usepackage{color}
\usepackage{bm}

\newcommand{\abs}[1]{\left|#1\right|}								% Betrag von #1

\begin{document}

%%%%%%%%%%%%%%%%%%%%%%%%%%%%%%%%%%%%%% AUTHORS %%%%%%%%%%%%%%%%%%%%%%%%%
\author{Peter Kroiss}
\author{Lode Pollet}
\affiliation{Department of Physics, Arnold Sommerfeld Center for Theoretical Physics and Center for NanoScience, University of Munich, Theresienstrasse 37, 80333 Munich, Germany}

%%%%%%%%%%%%%%%%%%%%%%%%%%%%%%%%%%%%%%%%%%%%%%%%%%%%%%%%%%%%%%%%%%%%%%%%%%%%%%
\title{Diagrammatic Monte Carlo study of quasi-two-dimensional Fermi-polarons}
%%%%%%%%%%%%%%%%%%%%%%%%%%%%%%%%%%%%%%%%%%%%%%%%%%%%%%%%%%%%%%%%%%%%%%%%%%%%%%

\date{\today}

\begin{abstract}
We apply a diagrammatic Monte Carlo method to the problem of an impurity interacting resonantly with a homogeneous Fermi bath for a quasi-two-dimensional setup. 
Notwithstanding the series divergence, we can show numerically that the three particle-hole diagrammatic contributions are not contributing significantly to the final answer, 
thus demonstrating a nearly perfect destructive interference of contributions in subspaces with higher-order particle-hole lines.
Consequently, for strong enough confinement in the third direction, the transition between the polaron and the molecule ground state is found to be in good agreement with the pure two-dimensional case and agrees very well with the one found by 
the wave-function approach in the two-particle-hole subspace.
\end{abstract}


\maketitle

%%%%%%%%%%%%%%%%%%%%%%%%%%%%%%%%%%%%%%%%%%%%%%%%%%%%%%%%%%%%%%%%%%%%%
\section{Introduction}
%%%%%%%%%%%%%%%%%%%%%%%%%%%%%%%%%%%%%%%%%%%%%%%%%%%%%%%%%%%%%%%%%%%%%

The Fermi-polaron problem consists of an impurity interacting resonantly with a noninteracting bath of fermionic atoms.
It constitutes an important limiting case of two-component Fermi mixtures with population imbalance:
If the concentration of one of the components is reduced, one can imagine that the system will eventually be dilute enough such that it can be described by independent single impurities. 
Although the Fermi-polaron problem may be believed to shed light on the phase diagram of imbalanced Fermi gases, the polaron to molecule transition is, in reality, precluded by phase separation (at zero temperature) 
into a fully polarized normal phase and an unpolarized superfluid phase\cite{parish2013,giorgini2008,bertaina2011}.

The Fermi-polaron problem was addressed in three dimensions
by Chevy via a variational ansatz in the single-particle-hole (1-ph) subspace\cite{chevy2006}. Later, Prokof'ev and Svistunov not only published improved
results for the polaron energies by means of the diagrammatic Monte Carlo method but also showed that a change in the ground-state wave function can favor a molecular state, consisting of the impurity and 
one bath particle\cite{prokofev2008A, prokofev2008B}. This molecular state has spin 0, whereas the polaron is a spin-1/2 quasiparticle.

At unitarity a dramatic cancellation of higher-order terms was seen, which was understood in Ref.~\onlinecite{Combescot}
from a nearly perfect destructive interference of contributions in subspaces with higher-order ph lines.
Thermodynamic quantities such as the effective mass, the residue, and the contact have been determined in Refs.~\onlinecite{punk2009,enss2011}.
The mass imbalance case was studied in Ref.~\onlinecite{Mathy2011}, where it was found that trimers (a bound state of the impurity with two majority atoms) can also form the ground state depending on the mass ratio.

With ultracold atoms~\cite{zwerger2008} a quasi-two-dimensional (quasi-2D)
geometry can be made by confining the dilute gas strongly in a plane by an external laser, which was done successfully in the experiment in Ref.~\onlinecite{koehl2012} for
fermionic atoms.
The quantum simulation of a 2D Fermi system may provide new insight in, e.g., high-$T_c$ superconductors. In order to keep an independent check on these experimental results, it is thus of prime importance to have better 
theoretical control over interacting 2D fermionic systems, including the 2D Fermi-polaron problem. 
Although it was initially believed that no polaron to molecule transition is possible~\cite{bruun2011} [due to an incorrect description of the Bose-Einstein-condensate (BEC) limit in lowest-order 
perturbation theory], it was shown that a transition can occur provided the molecules are dressed by ph fluctuations~\cite{parish2011}. Also trimers can be found for mass imbalance~\cite{parish2013}.
Given this initial controversy, a numerical calculation going beyond the lowest-order perturbation theory or the simplest variational ansatz is warranted in order to evaluate the smallness (or absence thereof) of the 
fluctuations beyond the first-order results. 

The diagrammatic Monte Carlo method (diagMC) is based on a sampling of high-dimensional Feynman diagram integrals allowing
extrapolation to infinite expansion order if the sign problem is not too severe. It can stochastically evaluate the integrals and the different topologies occurring in higher-order perturbation theory and thus provide 
an answer to the question posed above.
It has previously been applied to various fermionic problems such as the unitary Fermi gas~\cite{VanHoucke2012}, Anderson localization~\cite{Pollet2010}, the Hubbard model in the Fermi-liquid regime~\cite{Kozik2010}, 
and frustrated-spin systems~\cite{Kulagin2013, Kulagin2013bis}.
For the unitary Fermi gas and the three-dimensional (3D) Fermi-polaron problem it was found that the full many-body answer is very close to the first-order result given by a hole line on top of a $T$ matrix. For the problem of Anderson 
localization, it was found that the dynamical mean-field approach (exploiting the locality of the self-energy) was an excellent starting point. In all successful studies performed thus far, an underlying analytical 
understanding of the main physics allowed for an initial resummation which contained the dominant contributions, whereas the remaining fluctuations were rather small. One would hence expect that the success of diagMC 
for the Fermi-polaron can also be understood by identifying a (possibly emergent) small parameter. It remains an open question if the method can be successful when there are competing instabilities such as in the 
repulsive Hubbard model for low doping.

Recently, Vlietinck {\it et al.}~extended diagMC to the theoretical limit of pure 2D geometries~\cite{vlietinck2014}. For strong-enough interactions, they found growing fluctuations with expansion 
order that were claimed to be resummable and to result in a final answer close to the 1-ph result. In the following, we apply the diagrammatic Monte Carlo method to quasi-two-dimensional geometries for the Fermi-polaron problem
with equal mass as it would occur in a cold-gas experiment. Just as in Ref.~\onlinecite{vlietinck2014} we will see that the ground-state energy remains close to the 1-ph result and that the 2-ph result is almost 
quantitatively exact, with the remaining fluctuations being {\it very small} in ph-order. The small parameter that can be exploited in the diagrammatic approach is  the restricted phase space for the holes~\cite{Combescot}.

This paper is structured as follows:
Section \ref{sec:sec1} presents our basic ingredients for the quasi-two-dimensional diagrammatic Monte Carlo
simulation, focusing on the use of an appropriate $T$ matrix.
Section \ref{sec:sec2} introduces an alternative grouping technique designed to incorporate a new small parameter, the restricted volume for integration over hole momentum.
In Secs.~\ref{sec:sec3} and \ref{sec:sec4}, we present the results of our simulations.
Section \ref{sec:sec3} compares polaron and molecule energies for various interaction
strengths, demonstrating a polaron to molecule transition.
Next, Sec.~\ref{sec:sec4} checks the validity of the quasi-two-dimensional approach for various
confinement frequencies.
Finally, in Sec.~\ref{sec:sec5} we conclude by summarizing the main results.
%
%%%%%%%%%%%%%%%%%%%%
\section{Model}
\label{sec:sec1}
%%%%%%%%%%%%%%%%%%%%%
The Fermi-polaron problem consists of an impurity atom, labeled by $\downarrow$, interacting resonantly with majority atoms that form a non-interacting Fermi bath and are labeled by $\uparrow$.
We focus on equal masses $m$ of impurity and bath particles. 
The strength of the interaction is quantified by a bare coupling constant $g$. In order to extract model-independent, universal results, $g$ will be replaced by the two-particle scattering length $a$ in two dimensions, 
which has to be distinguished from $a_{\text{3D}}$, the three-dimensional scattering length. 
The Hamiltonian can be written as
\begin{equation}
\hat{H} = \sum_{\vec{k},\sigma} \epsilon_{k} \hat{c}_{\vec{k},\sigma}^{\dagger}
\hat{c}_{\vec{k},\sigma} + g \sum_{\vec{k},\vec{k'},\vec{q}}
\hat{c}_{\vec{k}+\vec{q},\uparrow}^{\dagger}\hat{c}_{\vec{k'}-\vec{q},\downarrow}^{\dagger}
\hat{c}_{\vec{k'},\downarrow} \hat{c}_{\vec{k},\uparrow}.
\end{equation}
$\hat{c}_{\vec{k},\sigma}$ and $\hat{c}^{\dagger}_{\vec{k},\sigma}$ are, respectively,
annihilation and creation operators of particles with spin $\sigma$ and momentum $\vec{k}$, and
$\epsilon_{k} = \frac{k^2}{2m}$ fixes the impurity and bath particle dispersions. We set $\hbar = 1$. 
The Fermi momentum and energy of the bath particles are denoted by $k_F$ and $E_F$, respectively, which we will use as units of inverse length and energy unless otherwise indicated.
We work at zero temperature.

In a realistic experiment, the 2D limit will be approached by confining the three-dimensional gas strongly along the $z$ axis by applying a  laser with  trapping frequency  $\omega_z$, thereby creating a tight harmonic
oscillator in the $z$ direction. We model this experimental setup by formally working in three dimensions but choosing $\omega_z$ high enough such that only the lowest harmonic oscillator state is populated.
The diagrammatic Monte Carlo method of Refs.~\onlinecite{prokofev2008A, prokofev2008B} can then straightforwardly be applied provided the momentum integrals are restricted to two dimensions and the relevant 
quasi-two-dimensional $T$ matrix is 
used, which simultaneously ensures that the zero-range limit of the interactions is taken.

The vacuum $T$ matrix $\Gamma_0$ is known analytically. Its low energy expression is derived in Refs.~\onlinecite{levinsen2012,pietilae2012} and reads
\begin{equation}
\Gamma_0(i\omega,\vec{k}) = \frac{\frac{4 \pi}{m}}{\frac{\sqrt{2\pi}l_z}{a_{\text{3D}}}  - \ln{(-\frac{\pi E}{B \omega_z})} + \ln{(2)} \frac{E}{\omega_z} },
\label{eq:gamma0}
\end{equation}
where $E  = i\omega + E_F + \mu_{\downarrow}^{0} - \frac{k^2}{4m}$, $B \approx 0.905$, $l_z = \sqrt{\frac{1}{m \omega_z}}$, and $\mu_{\downarrow}^{0}$ is an arbitrary parameter used for convergence reasons.
Note that this expression was already extended to imaginary frequencies.
The relationship between the two-dimensional scattering length and its three-dimensional correspondent is~\cite{levinsen2012,zwerger2008,petrov2001}
\begin{equation}
a = l_z \sqrt{\frac{\pi}{B}} \exp{\left(-\sqrt{\frac{\pi}{2}} \frac{l_z}{a_{\text{3D}}}\right)}.
\end{equation}
This defines the dimensionless interaction parameter $\eta = \ln(k_F a)$.
The two-body binding energy $E_B$ must be adjusted to reflect the quasi-two-dimensional nature of our model.
It is the solution of the following equation\cite{levinsen2012}:
\begin{equation}
\frac{l_z}{a_{\text{3D}}} = \mathcal{F}\left(\frac{E_B}{\omega_z}\right),
\end{equation}
where $\mathcal{F}$ is given by 
\begin{equation}
\mathcal{F}(x) = \int_0^\infty \frac{du} {\sqrt{4\pi u^3}} \left(1  -  \frac{\exp{(-x u)}}{\sqrt{[1-\exp{(-2u)}]/2u}}       \right) .
\end{equation}
The vacuum $T$ matrix can be linked with the in-medium $T$ matrix $\Gamma$ by~\cite{schmidt2012}
\begin{equation}
\begin{aligned}
\label{gammaomega}
\Gamma^{-1}(i\omega, \vec{k}) &= \Gamma_0^{-1}(i\omega, \vec{k}) \\
&+ \int_{\abs{\vec{k}} < k_F} \, \frac{d^2k } {(2 \pi)^2} \frac{1}{i\omega + E_F + \mu^0_\downarrow - \epsilon_{\vec{k}} - \epsilon_{\vec{k}+\vec{q}} }.
\end{aligned}
\end{equation}
This is the natural strong confinement expansion of Eq.~(10) of Ref.~\onlinecite{pietilae2012}.
For Monte Carlo sampling, the representation of $\Gamma(\tau,\vec{k})$ in imaginary time $\tau$ is required. It can be tabulated by Fourier transform
of  Eq.~(\ref{gammaomega}) prior to the main Monte Carlo run. In the pure 2D limit this transform is ambiguous because of the slow logarithmic decay with Matsubara frequency of $\Gamma(i\omega, \vec{k})$; see the 
middle term in the denominator of Eq.~(\ref{eq:gamma0}).
The regularization procedure induced by $\omega_z$ renders the Fourier transform
convergent by having $\Gamma \sim \frac{1}{\vert \omega \vert }$ for $\vert \omega \vert \gg \omega_z$ [see the last term in the denominator of Eq.~(\ref{eq:gamma0})] and can be physically motivated. The quasi-2D 
T-matrices were tested for consistency and accuracy by sampling the first-order polaron
diagram, a diagram that can be evaluated directly in $\omega$ space for the pure and quasi-2D setups.

The Monte Carlo algorithm we use is similar in spirit to the one developed by Prokof'ev and Svistunov~\cite{prokofev2008A, prokofev2008B} but differs in implementation
and update procedures. We have run extensive tests to ensure the correctness of both approaches.

%%%%%%%%%%%%%%%%%%%%
\section{Expansion technique}
\label{sec:sec2}
%%%%%%%%%%%%%%%%%%%%%
Arguably, the main bottlenecks in diagrammatic Monte Carlo are series convergence and the sign problem. There is no guarantee that the perturbative expansion in Feynman diagrams is a convergent series; 
in fact, some of the most famous theories in physics, such as quantum electrodynamics, are asymptotic~\cite{Dyson1952}.
Although the Dyson series for the Fermi-polaron in 3D at unitarity experiences diminishing fluctuations with increasing expansion order~\cite{prokofev2008A, prokofev2008B}, 
such cannot be assumed for fermionic many-body problems in general. The series is often nonmonotonous, showing increasing fluctuations~\cite{vlietinck2012, vlietinck2014}, and given the low expansion 
orders that can be reached (of the order of 12 in 3D~\cite{vlietinck2014} and 8 in 2D~\cite{vlietinck2012} for the Fermi-polaron problem), it is impossible to know the fate of the series convergence 
by inspecting order-by-order results. In such cases the best one can do is resort to resummation techniques provided the series is resummable. All resummation techniques that are strong enough to 
overcome the divergence of the series must then necessarily give the same result for infinite extrapolation order. Typically, Abelian resummation techniques were used in 
Refs.~\onlinecite{VanHoucke2012}, \onlinecite{vlietinck2014} and \onlinecite{vlietinck2012}, characterized by a 
very strong suppression of higher-order self-energy contributions, whereas weaker resummation methods did not yield a unique answer. 
Although the extrapolated results seemed to agree within the (small) error bars, this is at best a hint, and ultimately, only ``nature can provide the proof''~\cite{VanHoucke2012}.

\begin{figure}[tb]
\includegraphics[width=0.99\linewidth]{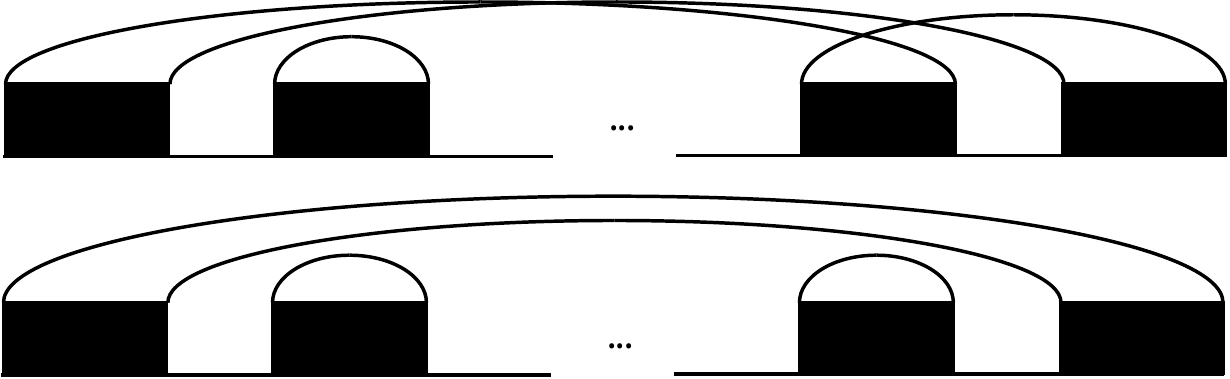}
\caption{\label{fig:diagram}
Examples of (top) ``exchange-hole'' and (bottom) ``direct-hole'' contributions to the 2-ph diagrams are shown. 
This demonstrates that every order $N>2$ has at least two diagrams counting as 2-ph.
}
\end{figure}

For a divergent series, regrouping terms is problematic and can result in any (unphysical) result. The standard approach groups the terms according to the number 
of T-matrices and sums these diagrams with the same coefficients.  We now discuss a second, physically motivated way of regrouping based on the arguments presented 
by Combescot and Giraud~\cite{Combescot}. These authors explained a remarkable cancellation of higher-order terms first seen in Refs.~\onlinecite{prokofev2008A, prokofev2008B} 
for the 3D Fermi-polaron at unitarity.
They argued that the subspace of $(n+1)$ ph pairs (and higher) can be decoupled from the subspace of $n$ ph pairs to a very good approximation because the 
summation over the particle lines dominates over the summation over the hole lines. The ground state in the single ph space is the Chevy ansatz and is 
already a very good approximation, whereas the ground state in the 2-ph subspace provides a small correction and so on. This provides a cascade of better, 
{\it variational} approximations. Diagrammatically, all contributions from the 1-ph spaces are contained in our lowest-order diagram. The 2-ph contributions 
can be identified~\cite{vlietinck2014} by taking all diagrams that have at most two particle and two hole lines at any moment in imaginary time. 
There are, in principle, an infinite number of them: although the two holes have only a direct contribution and an exchange contribution, the two particles can scatter arbitrarily.
This is illustrated in Fig.~\ref{fig:diagram}.
%%%%%%%%%%%%%%%%%%%%%%%%%%%%%%%%%%%%%%%%%%%%%%%%%
%%%%%%%%%%%%%%%%%%%%%%%%%%%%%%%%%%%%%%%%%%%%%%%%%
The sum of all these diagrams must yield the ground-state energy in the 2-ph subspace and agree with 
the wave-function ansatz, resulting in an upper bound to the true ground-state energy. This scheme suggests that one can regroup the diagrams in the number 
of ph lines and extrapolate the result. Within this subspace the diagrammatic expansion may still diverge, but if the resummation is inadequate, one 
may resort to other techniques (such as a brute force evaluation or variational Monte Carlo) to obtain the answer in this restricted subspace.

In the molecular sector, the 1-ph sector is already quantitatively accurate, as was demonstrated in Ref.~\onlinecite{punk2009} for a 3D polaron problem.
%
%%%%%%%%%%%%%%%%%%%%%%%%%%%%%%%%%%%%%%%%%%%%%%%%%
\begin{figure}[tb]
\includegraphics[width=0.99\linewidth]{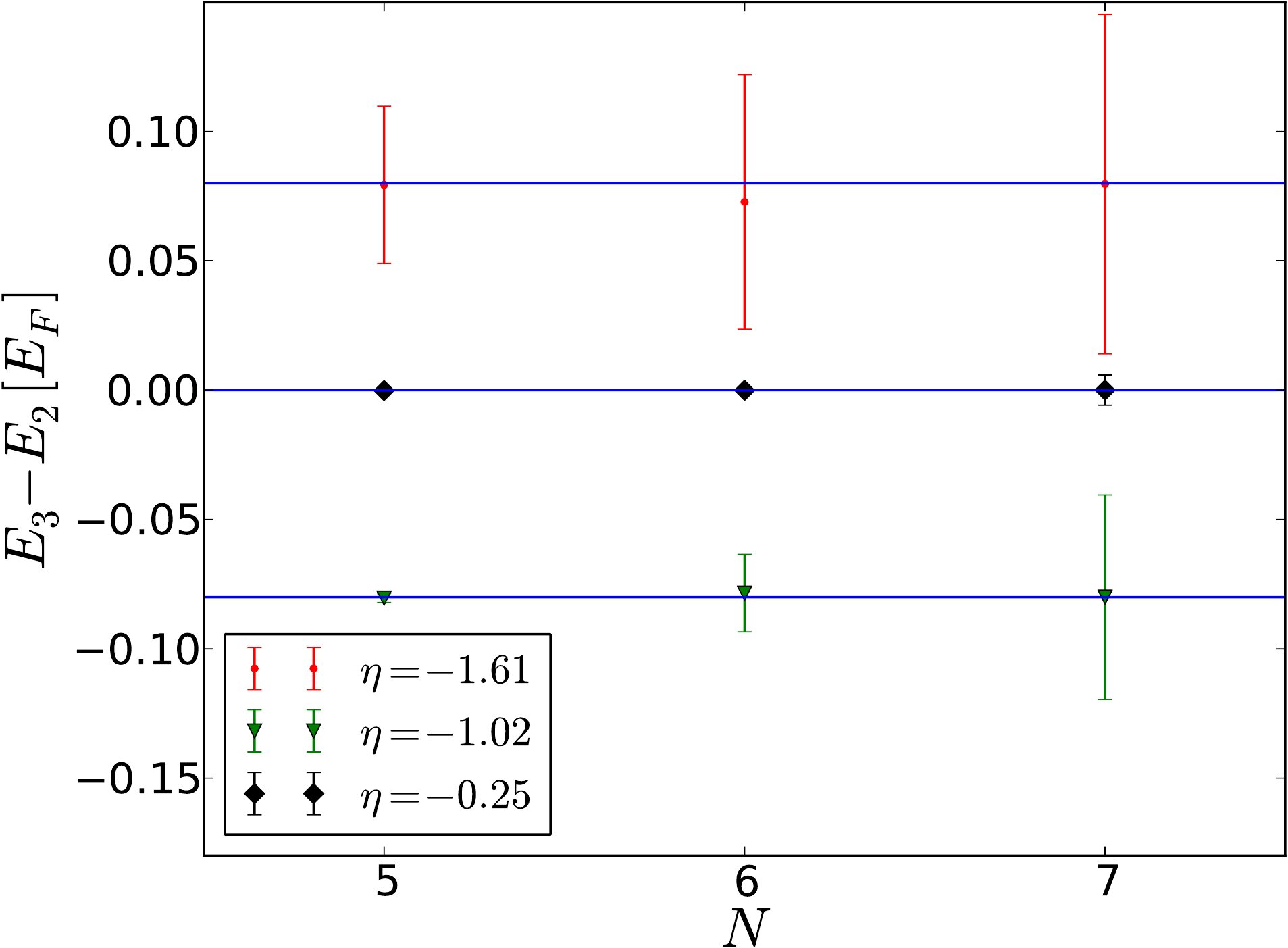}
\caption{\label{fig:resum_ph} (Color online)
Difference in energy between the 2-ph and 3-ph contributions as a function of the number of T-matrices $N$ in the Feynman diagrams.  Since this difference is essentially vanishing within the error bars, rapid 
convergence in the ph expansion order is seen.
Note that two of the data sets have been offset by $\pm 0.08 E_F$ for clarity.
}
\end{figure}
%%%%%%%%%%%%%%%%%%%%%%%%%%%%%%%%%%%%%%%%%%%%%%%%%

The regrouping works very well not only deep in the BCS phase but also in the vicinity of the polaron to molecule transition. 
As shown in Fig.~\ref{fig:resum_ph}, 
the difference between 2-ph and 3-ph contributions is vanishing on the order of the error bars, indicating that the 2-ph channel is already
sufficient for quantitatively precise calculations.
This holds for all accessible expansion orders, labeled by the number of T-matrices. We note that the 2-ph result itself has not converged up to the maximum expansion order, and its series 
is almost surely asymptotic. Nevertheless, resummation of the 2-ph series yields the same result as the wave-function technique, at least as long as the polaron is 
the true ground state. For nonzero momentum, this is no longer the case, and the polaron can decay. 
In the field theory this is signaled by a nonzero complex part of the particle self-energy (or finite width of the polaron peak in the spectral function). In such cases, the wave function is no longer 
variational, but the grouping in terms of the number of ph lines in diagMC is still a rapidly converging series (not shown).
We do not show results for 4-ph because the first contribution occurs for $T$ matrix expansion order 7, which gives us only a single point to this subspace.
In three dimensions, we observed vanishing contributions of 4-ph and 5-ph diagrams.

The above arguments show that using a bold code (with respect to the number of T-matrices as expansion order) is a questionable strategy: it mixes up the different 
contributions from different ph channels.
%
%%%%%%%%%%%%%%%%%%%%%%%%%%%%%%%%%%%
\section{Polaron-Molecule transition}
\label{sec:sec3}
\begin{figure}[tb]
\includegraphics[width=0.99\linewidth]{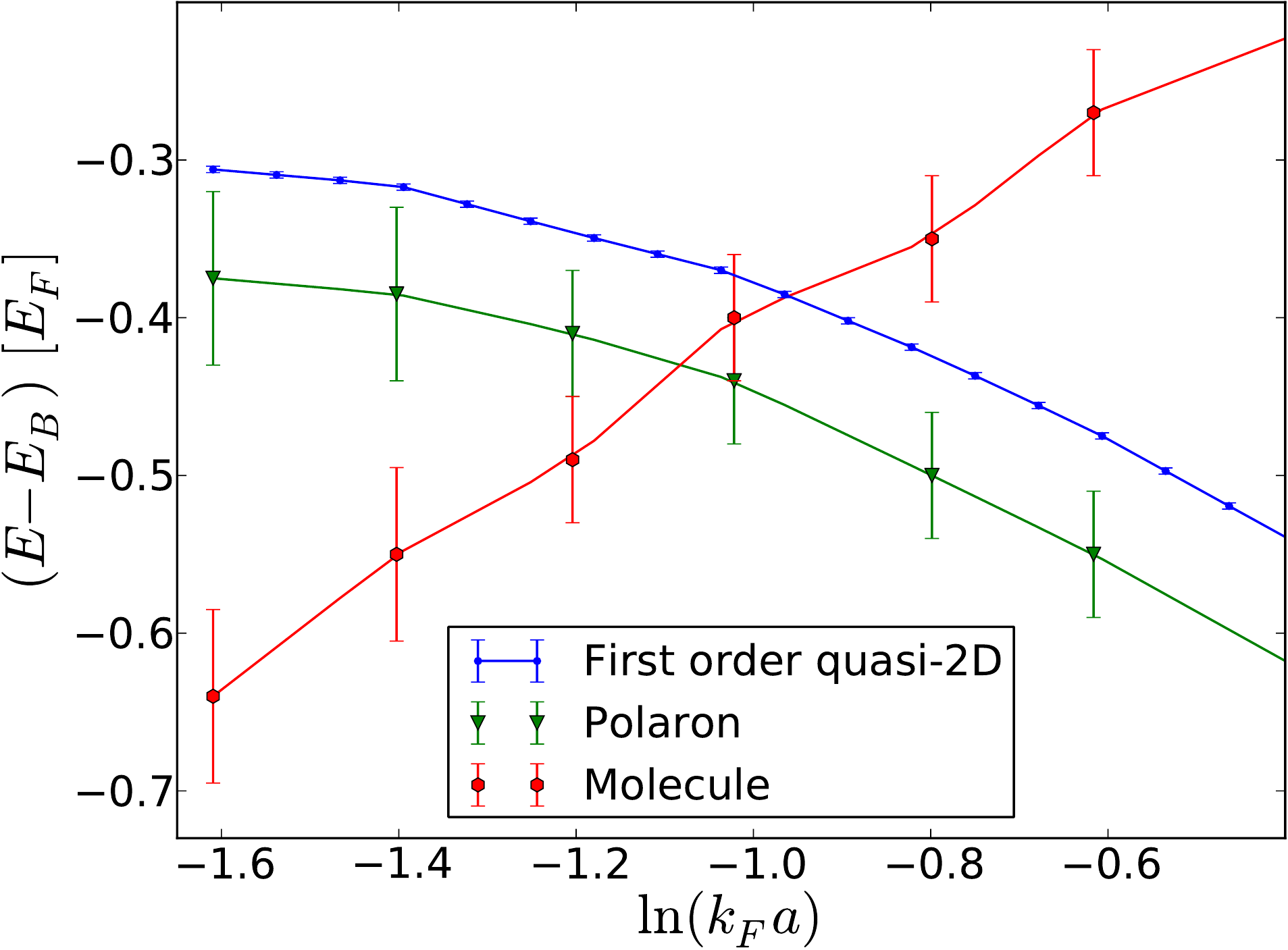}
\caption{\label{fig:crossover} (Color online)
Polaron and molecule energies as obtained by the diagrammatic Monte Carlo method.
For low values of the interaction parameter $\ln(k_F a)$, the molecule is the stable ground state, while
the polaron (green triangles) dominates in the weak-coupling regime $\ln(k_F a) \gtrsim -1 $. The 
first-order quasi-two-dimensional energy is also shown for comparison, from which we see that the  many-body modification is nearly independent of the interaction strength.
These data were produced for $\omega_z = 5000 E_F$.
}
\end{figure}

%%%%%%%%%%%%%%%%%%%%%%%%%%%%%%%%%%
Our results for the polaron and molecule energies in the quasi-two-dimensional geometry are shown in Fig.~\ref{fig:crossover}.
For weak two-body coupling $\ln(k_F a) \gtrsim -1$ the polaron state is the stable ground state, thus 
identifying the BCS regime in the limit $\ln(k_F a) \to +\infty$; for $\ln(k_F a) \lesssim -1$, the molecule becomes energetically
favorable and is referred to as the BEC regime in the limit $\ln(k_F a) \to -\infty$.
These curves were sampled at $\omega_z = 5000 E_F$, which is high enough to 
justify the assumption of only populating the lowest oscillator mode: $\frac{E_B}{\omega_z}$ is at most
1/70. The plot also shows the first-order contribution to the series, which is normally very close to the final result~\cite{vlietinck2014}. 
We used the standard approach with the number of T-matrices as the expansion parameter to perform the resummation, but the resummation method we used depends on the size of the binding energy $E_B$: 
For large $E_B$, it becomes necessary to use sharper Riesz resummations~\cite{prokofev2008B} on the reordered series. The error bars in Fig.~\ref{fig:crossover} 
appear to increase when one goes deeper in the BEC phase. This is, to a large extent, the result of the rapid increase in $E_B$, which is subtracted in the plot, whereas the calculations produce error bars on $E$ instead of $E-E_B$.
The pure two-dimensional first-order contribution $(E-E_B)$ curve
agrees with the quasi-two-dimensional one within the error bars.

We find the crossing point at $\ln(k_F a) = -1.1 \pm 0.2$, which is in good agreement with previous studies~\cite{vlietinck2014} and experiment~\cite{koehl2012}
and indistinguishable from the 2-ph result within our error bars, as could have been expected from the previous discussion.

For molecular energies, the series is alternating and can be well resummed with Riesz techniques.
On the BCS side of the transition, the molecule is not stable any longer, which leads to a breakdown of the Monte Carlo estimators for energies above $-E_F$.

In principle, our scheme could be used to calculate the effective mass and contact coefficient of the system.
However, as the error bar is on the scale of the difference between the first-order result and the extrapolated result,
it is not reasonable to extract quantities depending on derivatives from our data. However, given the quantitative accurateness of the 2-ph result, 
precise estimates of ground-state quantities (such as the contact) can be obtained within the 2-ph subspace.
%
%%%%%%%%%%%%%%%%%%%%%%%%%%%%%%%%%%%%%%%%%%%%%%%%%
%%%%%%%%%%%%%%%%%%%%%%%%%%%%%%%%%%%%%%%%%%%%%%%%%
%
%%%%%%%%%%%%%%%%%%%%%%%%%%%%%%%%%%%
\section{Validity of the quasi-two-dimensional approach}
\label{sec:sec4}
\begin{figure}[tb]
\includegraphics[width=0.99\linewidth]{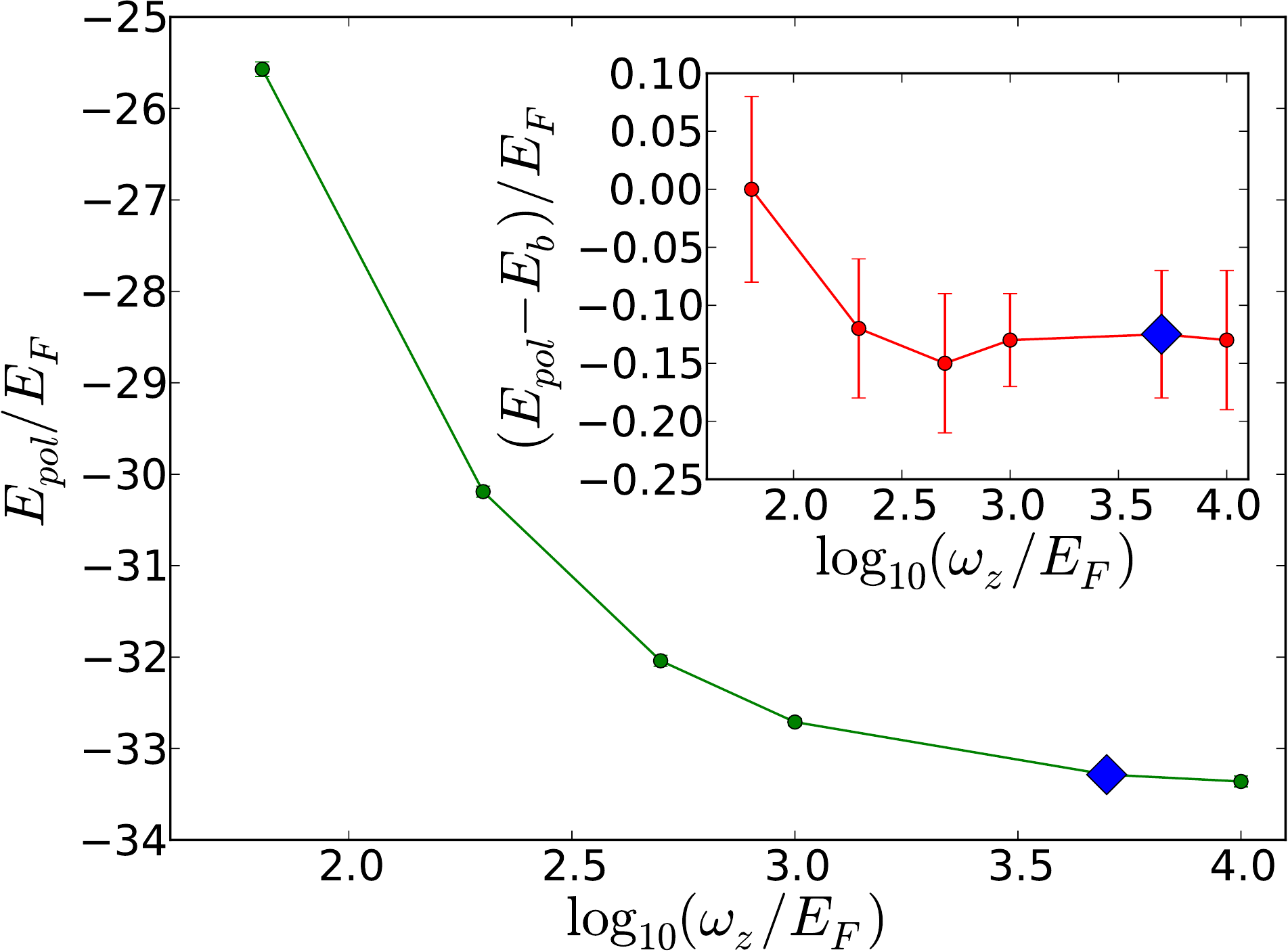}
\caption{\label{fig:confinement} (Color online) 
The influence of $\omega_z$ on both polaron energy and binding energy is demonstrated.
Using $\omega_z = 5000 E_F$, the results are well saturated, justifying the assumption of neglecting transitions between the lowest and higher harmonic oscillator levels. The data in the plot were 
measured at $\ln(k_F a) = -1.4$. Blue diamonds mark the value of $\omega_z$ we use in our simulations. Riesz resummation was applied to the bare data.
}
\end{figure}

%%%%%%%%%%%%%%%%%%%%%%%%%%%%%%%%%%
The approach to the 2D limit used in this paper consists of using a strong harmonic confinement in the $z$ direction
and assuming that only the lowest harmonic oscillator is populated; that is, we neglect transitions between different harmonic oscillator levels.
For strong enough confinement this approach is physically justified, and the 2D limit can be found by extrapolating results obtained for different $\omega_z$.
In order to check the validity of this quasi-2D approach and the corresponding quasi-2D T-matrices, we compare 
polaron energies for several values of the confinement frequency $\omega_z$ in Fig.~\ref{fig:confinement}.
It is remarkable that even the loosest confinement $\omega_z = 2 E_B$ (which clearly violates the condition
of populating only one mode of the oscillator in the $z$ direction) shows good agreement for $E_{\text{pol}}-E_B$.
One would expect that high values of $\omega_z$ are necessary to reproduce the pure 2D limit because of the logarithmic dependence of the energy scale.
Indeed, we see that $\omega_z = 5000 E_F$ is high enough to observe the polaron-molecule crossover in this limit
(it will be insufficient deep in the BEC phase for the polaron energy, however, because it has to be kept in relation to the binding energy to ensure exclusive population of the lowest mode).
Lower values of $\omega_z$ may be acceptable too if $E-E_B$ is calculated.
In the polaron experiment of Ref.~\onlinecite{koehl2012}, a confinement frequency of $\omega_z \approx 7.9 E_F$ was used.
%
%%%%%%%%%%%%%%%%%%%%%%%%%%%%%%%%%%%%%%%%%%%%%%%%%
%%%%%%%%%%%%%%%%%%%%%%%%%%%%%%%%%%%%%%%%%%%%%%%%%
%
%%%%%%%%%%%%%%%%%%%%%%%%%%%%%%%%%%%
\section{Conclusion}
\label{sec:sec5}
%%%%%%%%%%%%%%%%%%%%%%%%%%%%%%%%%%
We have applied the diagrammatic Monte Carlo method to a quasi-2D Fermi-polaron problem.
In any realistic cold-gas experiment investigating the 2D Fermi-polaron problem, a laser with a strong trapping frequency in the $z$ direction has to be applied, which we took into account in our model. The validity of
this approach was checked by comparing results for different trapping frequencies and the pure 2D limit, showing good agreement for the confinement we used.

The resulting transition point between the polaron and molecular ground states is shifted with respect to the variational first-order calculations but is in very good agreement with variational results in the 2-ph subspace. 
Our Monte Carlo results have shown that the difference between 2-ph and 3-ph contributions is vanishing within the error bars, and this holds order per order in the Feynman expansion using the $T$ matrix as the expansion
parameter.  We therefore suggest computing the 2-ph contributions by using the wave-function approach and switching to diagMC for the computation of corrections to the 2-ph contributions.
The number of hole lines can still be used as the expansion parameter for polaron problems at finite momentum, where the wave-function approach is no longer variational. It is the restricted phase space for hole 
excitations~\cite{Combescot} that enables this.

%%%%%%%%%%%%%%%%%%%%%%%%%%%%%%%%%%%%%%%%%%%%%%%%
We are grateful to M. Bauer, J. Levinsen, M. Parish, N. Prokof'ev, R. Schmidt, B. Svistunov, K. Van Houcke, and W. Zwerger for valuable discussions. This work is supported
by the Excellence Cluster NIM, FP7/Marie-Curie Grant No. 321918 (FDIAGMC), and FP7/ERC Starting Grant No. 306897 (QUSIMGAS).

Note added: Recently, a similar publication by Vlietinck {\it et al.}~\cite{vlietinck2014} became public.
Results agree where applicable.

\end{document}